\documentclass[twocolumn,showpacs,preprintnumbers,amsmath,amssymb]{revtex4}

\usepackage{graphicx}
\usepackage{epsfig}
\usepackage{xcolor}
\usepackage{dcolumn}

\usepackage{longtable}
\usepackage{longtable}

\begin{document}


\title{Entropies, level-density parameters and fission probabilities along the triaxially- and
axially-symmetric fission paths in  $^{296}$Lv}


\author{A. Rahmatinejad}
\affiliation{Joint Institute for Nuclear Research, Dubna, 141980, Russia}
\author{T. M. Shneidman}
\altaffiliation[Also at ]{Kazan Federal University, Kazan 420008, Russia}
\affiliation{Joint Institute for Nuclear Research, Dubna, 141980, Russia}
\author{G. G. Adamian}
\affiliation{Joint Institute for Nuclear Research, Dubna, 141980, Russia}
\author{N. V. Antonenko}
\affiliation{Joint Institute for Nuclear Research, Dubna, 141980, Russia}

\author{P.~Jachimowicz}
 \affiliation{Institute of Physics,
University of Zielona G\'{o}ra, Szafrana 4a, 65516 Zielona G\'{o}ra, Poland}

\author{M.~Kowal} \email{michal.kowal@ncbj.gov.pl}
\affiliation{National Centre for Nuclear Research, Pasteura 7, 02-093 Warsaw, Poland}
\date{\today}

\begin{abstract}
We employ a statistical approach to investigate the influence of axial asymmetry on the nuclear level density and entropy along the fission pathways of a superheavy nucleus, specifically focusing on the $^{296}$Lv isotope. These pathways are determined within multidimensional deformation spaces. Our analysis reveals a significant impact of triaxiality on entropy. Additionally, suppressing shell effects can alter the fission scenario depending on the available excitation energy. We derive the deformation-dependent level density parameter, which plays a crucial role in estimating the survival probability of a superheavy nucleus. Furthermore, we utilize a set of master equations to obtain the time-dependent fission probabilities and calculate the ratio of decay probabilities for both axial and triaxial paths
\end{abstract}

\maketitle

\section{\label{sec:level1}Introduction}

Recent breakthroughs in synthesizing novel superheavy nuclei, as documented in the works of Oganessian et al. \cite{Og1, Oganessian1, Oganessian2}, have resulted in substantial advancements. Building upon these achievements, forthcoming experiments are planned to extend the frontiers of heavy element production, capitalizing on the cutting-edge "superheavy factory" facility recently established at JINR (Dubna).

The successful production of superheavy nuclei relies heavily on their ability to resist fission. This resistance is crucial in determining the survival probability of a hot compound nucleus formed in complete fusion reactions. The competition between neutron emission and fission plays a significant role in shaping the outcome of the decay process, and it depends on factors such as the level densities and the potential energy surface (PES) topology. These characteristics mainly predestine the formation of evaporation residue nuclei.
The role of entropy in this process is less studied and was not frequently taken up in studies.
The critical advantage of entropy is its ability to encode information about the density of states across diverse potential energy landscapes simultaneously.
However, the concept of entropy becomes more nuanced and challenging to define precisely.
This is primarily due to the following reasons:

\begin{itemize}
  \item Finite Size: Unlike macroscopic systems, which typically consist of many particles, heavy nuclei are relatively small and finite systems. As a result, the statistical behavior and thermodynamic quantities, including entropy, might exhibit deviations from the behavior predicted by classical statistical mechanics.
  \item Quantum Mechanical Nature: Nuclei are governed by quantum mechanics, which introduces inherent uncertainties and restrictions on the states available to individual particles. The discrete energy levels and quantum correlations make entropy more intricate than classical systems.
  \item  Lack of Equilibrium: Determination of the entropy in a nucleus requires achieving thermal equilibrium, which is challenging for isolated individual nuclei. Unlike macroscopic systems that readily reach equilibrium through interactions with a heat bath, individual nuclei are not typically in thermal contact with a large reservoir, making the application of equilibrium statistical mechanics less straightforward.

\end{itemize}
Nevertheless,  various statistical approaches and theoretical models have been developed to estimate and describe the entropy of heavy nuclei. These methods incorporate the concepts of nuclear level densities, statistical ensembles, and thermodynamic considerations to provide insights into the statistical behavior and thermodynamic properties of atomic systems. So, the idea of entropy is still in use.

Calculating level densities, we should determine the eigenvalues and their degeneracy of the nuclear Hamiltonian and count the number of states within a specific energy interval of interest. Due to the exponential increase in the total number of states with excitation energy above a few MeV, statistical methods are employed to tackle the problem. Several sophisticated combinatorial methods, such as those described in Refs.~\cite{hilary,hilary2}, have been developed, where factors like parity, angular momentum, pairing correlations, and collective enhancements are explicitly accounted for using the Gogny interaction. Additionally, a general and exact scheme for calculating particle-hole level densities while considering the Pauli exclusion principle is presented in Ref. \cite{Blin}. The role of microscopic level densities in fission and their influence on nuclear shape evolution with excitation energy are discussed in Refs.~\cite{Aberg1Aberg3Aberg4}. Relativistic mean-field theory calculates atomic level densities in Ref. \cite{pomorska1}.
In contrast, a self-consistent mean-field approach is utilized to determine single-particle level densities at various temperatures. The level density parameters are calculated using the Yukawa-folded potential in Ref. \cite{pomorska2pomorska3}. Spin- and parity-dependent nuclear level densities obtained with the moment method in the proton-neutron formalism are presented in Ref. \cite{Senkov}. Finally, a direct microscopic calculation of nuclear level densities with the shell model Monte-Carlo approach is discussed in Ref. \cite{Alhassid2015}. Certain approximations and assumptions are still necessary for practical applications, including corrections for superfluidity effects and collective rotational and vibrational enhancements.

The well-established Fermi-gas model can effectively describe the transition from a paired system to a system of non-interacting fermions as the nuclear system goes from low to higher energies. This phenomenological model accounts for the pairing effect by introducing a constant back-shift parameter $\Delta$ of the excitation energy. In the Fermi-gas model, the average level-density parameter, which relates the excitation energy to the nuclear temperature, is often assumed to have a linear dependence on the mass number $A$ \cite{Sokolov1990}. However, the level-density parameter exhibits energy dependence and gradually approaches an asymptotic value as the energy increases beyond the neutron separation energy. To incorporate the energy and shell correction dependencies into the level density parameter, a phenomenological expression was introduced in Ref.~\cite{Ignatyuk1975}.

Our recent findings \cite {Rahmatinejad2021} have demonstrated that the level density parameter $a$, which plays a crucial role in estimating survival probabilities, strongly depends on the available excitation energy, particularly at low energies. Given its appearance as an exponential function in the decay rates, even minor variations in this parameter can profoundly impact the estimated survival probabilities of superheavy nuclei. In this study, we will derive this parameter as a deformation function, providing a more accurate and comprehensive description.

The primary objective of this article is to combine advanced techniques, including the multidimensional manifold of deformations for fission trajectories, with a statistical approach for computing nuclear level density and entropy and calculating the level density parameter along fission paths. We discuss the calculations of two extreme cases, axial and nonaxial fission pathways for the $^{296}$Lv nucleus.
We do not consider deformations that lead to mass-asymmetric fission, such as octupole deformations, because they become significant  beyond the scope of the first fission barrier only that is the primary interest of our study \cite{Abusara2012,Scamps2015,Zhao2016}.

It is important to note that the level density at any point along specific fission paths is not the same as at the ground state (GS). This is primarily due to two factors. Firstly, the excitation energy available at any point of the fission paths is reduced by the difference in deformation energy between this point and the GS. Secondly, the distribution of single-particle levels is altered due to the change in the shape of the compound nucleus. Thus, identifying the correct paths that govern
the fission process is essential.

\section{\label{sec:level12} Method of Calculation}

We adopt a two-step methodology in our study. Firstly, we construct a 5-dimensional PES using the macroscopic-microscopic (MM) method. This PES allows us to identify suitable axial and nonaxial fission paths by considering the direction with the most substantial gradient drop in all deformation variables. This minimization procedure ensures movement along the adiabatic or lowest energy path. Such a scenario lacks distinct configuration or structural effects, which could be crucial in fission at very low excitation energies. Therefore, we assume that the adiabatic process predominantly governs the system's behavior and properties, with minimal influence from specific structural features. The presumption that the fission process from the ground state to the first saddle point, and slightly beyond, is adiabatic, suggesting an extremely slow progression (ground state to saddle: 1,000,000 zs; saddle to scission: 10 - 100 zs), is well-founded, see for example in \cite{vandenbosch1973nuclear,duRietz2011predominant,morjean2008fission,morjean2013long, bulgac2016, andreyev2018}
and in the citations therein.
In the second step, we employ a statistical approach to determine the density of states. Based on this information, we calculate the entropy and relevant parameters along the fission paths.

\subsection{\label{sec:lev} Potential Energy Surface}

The PES are calculated using the MM  method. In this approach, the microscopic energy is determined by employing the Strutinski shell and pairing correction method \cite{STRUT67}, which involves diagonalizing the deformed Woods-Saxon potential \cite{WS} to obtain single-particle levels. We consider $n_{p}=450$ lowest proton levels and $n_{n}=550$ lowest neutron levels from the $N_{max}=19$ lowest shells of the harmonic oscillator in the diagonalization process. The standard values of $\hbar\omega_{0}=41/A^{1/3}$ MeV for the oscillator energy and $\gamma=1.2 \hbar\omega_{0}$ for the Strutinski smearing parameter are used. Additionally, a six-order correction polynomial is applied to calculate the shell correction.

For the macroscopic component, we employ the Yukawa plus exponential model \cite{KN} with parameters specified in Ref.~\cite{WSpar}. The deformation-dependent Coulomb and surface energies are integrated using a 64-point Gaussian quadrature method.

To construct the  PES,
a five-dimensional deformation space is utilized, which includes an expansion of the nuclear radius:
\begin{eqnarray} \label{sp1}
R(\vartheta ,\varphi) = c R_0\{
1 & + & \beta_{2 0} {\rm Y}_{2 0} + {\beta_{2 2} \over {\sqrt{2}}} \lbrack {\rm Y}_{2 2} + {\rm Y}_{2 -2} \rbrack  \nonumber \\
  & + & \beta_{4 0} {\rm Y}_{4 0} +  \beta_{6 0} {\rm Y}_{6 0} +
 \beta_{8 0} {\rm Y}_{8 0} \} ,
\end{eqnarray}
where the quadrupole non-axial deformation parameter $\beta_{2 2}$ is explicitly included.
For each nucleus, we generate the following 5D grid of deformations:
\begin{eqnarray} \label{sp1grid}
\beta_{2 0} & = & \phantom {-} 0.00 \; (0.05)  \; 0.60, \nonumber \\
\beta_{2 2} & = & \phantom {-} 0.00 \; (0.05)  \; 0.45, \nonumber \\
\beta_{4 0} & = &           -  0.20 \; (0.05)  \; 0.20,           \\
\beta_{6 0} & = &           -  0.10 \; (0.05)  \; 0.10, \nonumber \\
\beta_{8 0} & = &           -  0.10 \; (0.05)  \; 0.10. \nonumber
\end{eqnarray}
A grid of $29,250$ points (nuclear shapes) is employed, with the grid steps specified in parentheses. This primary grid, labeled as (\ref{sp1grid}), is extended through fivefold interpolating in all directions. As a result, an interpolated energy grid consisting of over 50 million points is obtained (see Refs. \cite{Jachimowicz2017_II, Jachimowicz2017_III} for more details).
All the parameters are specified in the last tables of Ref. \cite{table2021}.

\subsection{\label{sec:lev1} Level-density parameters}

Based on the superfluid formalism \cite{Decowski1968,origin1} and  an assumption of thermal equilibrium between neutron and proton subsystems, the excitation energies $(U=U_{Z}+U_{N})$, entropies $(S=S_{Z}+S_{N})$ and intrinsic level densities $\rho$ are calculated at each  temperature $T$ as
\begin{equation} \label{eq6}
E_{N(Z)}(T)=2\sum_k\varepsilon_k n^{N(Z)}_{\Delta,k}-\frac{\Delta^{2}_{N(Z)}}{G_{N(Z)}},
\end{equation}
\begin{equation} \label{eq7}
U_{N(Z)}(T)=E_{N(Z)}(T)-E_{N(Z)}(0),
\end{equation}

\begin{eqnarray} \label{eq8}
S_{N(Z)}(T)=\sum_{k}\left\{\ln\left[1+\exp{\left(-\frac{E^{N(Z)}_{k}}{T}\right)}\right]\right.\nonumber \\
\left.
+\frac{  E^{N(Z)}_{k}}{T\left[1+\exp{\left(\frac{E^{N(Z)}_{k}}{T}\right)}\right]}\right\},
\end{eqnarray}

\begin{equation} \label{eq9}
\rho=\frac{\exp{(S)}}{(2\pi)^{\frac{3}{2}}\sqrt{D}}.
\end{equation}
The determinant of the matrix composed of the second derivatives of the entropy concerning $1/T$ and $\mu=\lambda/T$ is denoted as $D$.
In the equations above, the occupation probabilities
\begin{equation} \label{eq5}
 n_{\Delta,k}^{N(Z)}(T)=\frac{1}{2}\left(1-\frac{\varepsilon^{N(Z)}_{k}-\lambda_{N}}{E^{N(Z)}_{k}}\tanh\frac{E^{N(Z)}_{k}}{2T}\right)
\end{equation}
and the quasiparticle energies $E^{N(Z)}_{k}=\sqrt{(\varepsilon^{N(Z)}_{k}-\lambda_{N(Z)})^2+\Delta_{N(Z)}^2}$ are calculated using the single-particle energies ($\varepsilon^{N(Z)}_{k}$) of the Woods-Saxon potential. The constants of the pairing interaction for neutrons ($G_{N}$) and protons ($G_{Z}$) are adjusted to obtain correct  GS  pairing gaps ($\Delta_{N}$ and $\Delta_{Z}$) in the MM method with using the Bardeen-Cooper-Schrieffer (BCS) equations:
\begin{equation} \label{eq2}
N = 2\sum_k n^{N}_{\Delta,k}(T),
\end{equation}
\begin{equation} \label{eq3}
\frac{2}{G_{N}}=\sum_k\frac{1}{E^{N}_{k}}\tanh\frac{E^{N}_{k}}{2T}
\end{equation}
for neutrons and
\begin{equation} \label{eq4}
Z = 2\sum_k n^{Z}_{\Delta,k}(T),
\end{equation}
\begin{equation} \label{eq5b}
\frac{2}{G_{Z}}=\sum_k\frac{1}{E^{Z}_{k}}\tanh\frac{E^{Z}_{k}}{2T}
\end{equation}
for protons at zero temperature.
Using the values of pairing constants obtained, the pairing gaps $\Delta_{N(Z)}$ and chemical potentials $\lambda_{N(Z)}$
are determined by solving Eqs.(\ref{eq2})--(\ref{eq5b}) at given temperatures.
We repeated the calculations using the single-particle level energies obtained with the Woods-Saxon potential at each given set of deformations ($\beta=\beta_{20},\beta_{22}$) along the fission path.

Above the critical temperature ($T_{cr}$), the pairing gap disappears, and all thermodynamic quantities revert to those of a noninteracting Fermi system. Generally, a more significant density of states close to the Fermi surface at the saddle point (SP) leads to a more considerable pairing correlation and, as a result, to a more significant critical temperature than the GS. In our calculations in the superheavy mass region, the critical temperatures for neutrons and protons are about $0.42$ MeV at the GS and $0.52$ MeV at the SP. The corresponding total excitation energies are approximately $U_{cr}\approx 5.14$ MeV at the GS and $U_{cr}\approx 11.27$ MeV at the SP.

Fitting the calculated values of the intrinsic level density with the back-shifted Fermi gas expression
\begin{equation} \label{eq10}
\rho_{FG}(U)=\frac{\sqrt{\pi}}{12 a^{\frac{1}{4}}(U-\Delta)^{\frac{5}{4}}}\exp({2\sqrt{a(U-\Delta)}})
\end{equation}
we obtain the level density parameter $a(U)$ as a function of excitation energy.
In the calculations, the energy back-shifts are taken as $\Delta=12/\sqrt{A}$, 0, and $-12/\sqrt{A}$ MeV for even-even, odd, and odd-odd isotopes, respectively.

At finite temperatures, the microscopic corrections to energy are replaced by microscopic corrections to the free energy $\delta{F}(T)=\delta{F}_{shell}(T)+\delta{F}_{pair}(T)$. Hence, the pairing and shell corrections to entropy and the energy are required at each temperature. In general, the temperature-dependent energy and entropy are defined as
\begin{equation} \label{eq13b}
E(T)=2\int_{-\infty}^{\infty} n(T,\lambda) \varepsilon g(\varepsilon) d\varepsilon
\end{equation}
and
\begin{eqnarray} \label{eq20}
S(T)=-\int_{-\infty}^{+\infty}[(1-n(T,\lambda))\log(1-n(T,\lambda))+\nonumber \\
n(T,\lambda) \log n(T,\lambda)] g(\varepsilon)d\varepsilon,
\end{eqnarray}
respectively. Here, $g(\varepsilon)$ is the density of the single-particle levels and temperature-dependent occupation probability is taken as $n(T,\lambda)=1/(1+e^{(\varepsilon-\lambda)/T})$ \cite{Ivanyuk2018}.

To obtain the shell correction at finite temperature $\delta{F}_{shell}=E(T)-\tilde{E}(T)-T\left[S(T)-\tilde{S}(T)\right]$, we calculate the energy $E(T)$ and entropy $S(T)$ for the discrete spectrum by taking $g(\varepsilon)=\sum_{k}\delta(\varepsilon-\varepsilon_{k})$ in (\ref{eq13b},\ref{eq20}). The energy $\tilde{E}(T)$ and entropy $\tilde{S}(T)$ for the smoothed spectrum are calculated with occupation number taken as $n(T,\tilde{\lambda})=1/(1+e^{(\varepsilon-\tilde{\lambda})/T})$ and
the smooth density $\tilde{g}(\varepsilon)$ of the single-particle levels as
\begin{equation} \label{eq12a}
\tilde{g}(\varepsilon)=\frac{1}{\gamma \sqrt{\pi}}\sum_{k=1}e^{-x^{2}}\sum_{n=0,2,...}^{6}c_{n}H_{n}(x),
\end{equation}
where  $x=(\varepsilon-\varepsilon_{k})/\gamma$, $c_{0}=1$, and $c_{n+2}=-c_{n}/(n+2)$ \cite{STRUT67}.
The chemical potentials ($\lambda, \tilde{\lambda}$) are defined by the particle number conservation condition
\begin{equation} \label{eq13c}
N(Z)=2\int_{-\infty}^{+\infty}n(T,\lambda) g(\varepsilon)=2\int_{-\infty}^{+\infty}n(T,\tilde{\lambda}) \tilde{g}(\varepsilon) d\varepsilon.
\end{equation}

The temperature-dependent pairing correction is calculated as $\delta{F}_{pair}=E_{pair}-\tilde{E}_{pair}-T\left[S_{pair}-\tilde{S}_{pair}\right]$.
The pairing energy, $E_{pair}(T)=E^{BCS}(T)-E^{BCS}_{\Delta=0}(T)$, corresponding to the real single-particle level distribution  is calculated using
\begin{equation} \label{eq13d}
E^{BCS}_{N(Z)}(T)=2\sum_k\varepsilon_k n^{N(Z)}_{\Delta,k}-\frac{\Delta^{2}_{N(Z)}}{G_{N(Z)}}-G\sum_{k}\left(n_{\Delta,k}^{N(Z)}\right)^2,
\end{equation}
where $n_{\Delta,k}^{N(Z)}$ is defined in Eq.~\eqref{eq5}.
Note that the last term of \eqref{eq13d} is often neglected in the calculation of excitation energy under the standard assumption of the BCS model \cite{Bolsterli1972}.
However, it should be considered in the microscopic corrections because of the different pairing constants used in the discrete and smoothed spectra calculations.

The pairing energy corresponding to the smooth single-particle level distribution $\tilde{E}_{pair}(T)=\tilde{E}^{BCS}(T)-\tilde{E}_{\Delta=0}^{BCS}(T)$ is calculated using
\begin{eqnarray} \label{eq14}
\tilde{E}^{BCS}_{N(Z)}(T)=2\int_{-\infty}^{+\infty}\varepsilon \tilde{g}(\varepsilon)\tilde{n}(T)d\varepsilon -
\frac{\tilde{\Delta}_{N(Z)}^{2}}{\tilde{G}_{N(Z)}}\nonumber \\
-\tilde{G}\int_{-\infty}^{+\infty}\tilde{g}(\varepsilon)\tilde{n}^{2}(T)d\varepsilon.
\end{eqnarray}
Here,
\begin{eqnarray} \label{eq18b}
\tilde{n}(T)=\frac{1}{2}\left(1-\frac{\varepsilon-\tilde{\lambda}_{N(Z)}}{\tilde{E}_{N(Z)}}\tanh\frac{\beta \tilde{E}^{N(Z)}}{2}\right)
\end{eqnarray}
is the occupation probability
and  $\tilde{E}^{N(Z)}=\sqrt{(\varepsilon-\tilde{\lambda}_{N(Z)})^2+\tilde{\Delta}_{Z(N)}^2}$ is the quasiparticle energy.
The chemical potentials $\tilde{\lambda}_{Z(N)}$ and pairing gaps $\tilde{\Delta}_{N(Z)}$ for smoothed spectrum are calculated by solving analogs of BCS equations:
\begin{equation} \label{eq15}
\frac{2}{\tilde{G}_{N(Z)}}=\int_{-\infty}^{+\infty}\frac{1}{\tilde{E}^{N(Z)}}\tanh\left(\frac{\beta \tilde{E}^{N(Z)}}{2}\right)\tilde{g}(\varepsilon)d\varepsilon,
\end{equation}
\begin{equation} \label{eq16}
N(Z) =\int_{-\infty}^{+\infty} \left(1-\frac{\varepsilon-\tilde{\lambda}_{N(Z)}}{\tilde{E}^{Z(N)}}\tanh\frac{\beta \tilde{E}^{N(Z)}}{2}\right)\tilde{g}(\varepsilon)d\varepsilon.
\end{equation}

The pairing constant in Eq.~\eqref{eq15} is adjusted to reproduce the smoothed pairing energy at $T=0$, which is obtained with the assumption of a constant density of pairs near the Fermi level \cite{table2021}.

The pairing correction to entropy is calculated similarly.
The average entropy $\tilde{S}_{N(Z)}^{BCS}(T)$ is obtained by replacing the sum over discrete states $k$ in Eq.\eqref{eq8} by integrals with the average density of states defined in Eq.~\eqref{eq12a} and replacing the quasiparticle energy $E^{N(Z)}_{k}$ by its smooth analog $\tilde{E}^{N(Z)}$.

\section{\label{sec:level13} Results and discussion}

\begin{figure}[h]
\begin{center}
\includegraphics[width=0.5\textwidth] {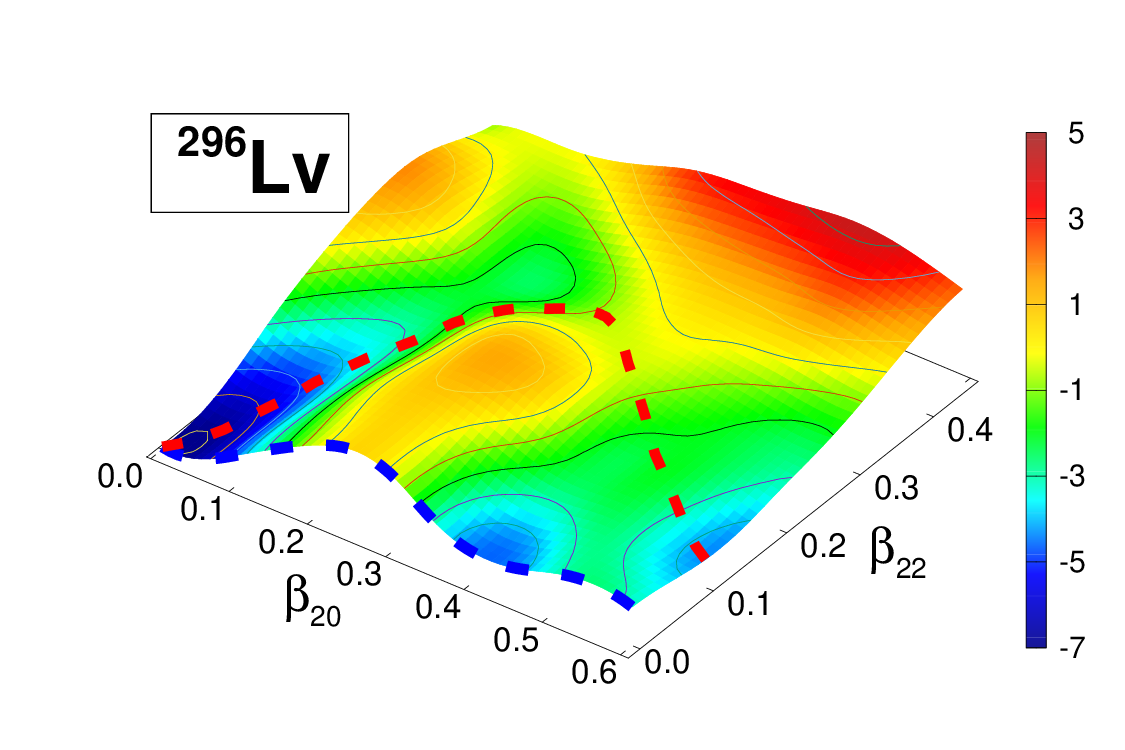}
\caption {The potential energy landscape of $^{296}$Lv is projected onto the $(\beta_{20},\beta_{22})$ plane. The dashed line represents the potential fission path from the  GS  to the  SP. The energy scale is measured in MeV and is calculated relative to the macroscopic energy at the spherical shape.}
\label{PESLv}
\end{center}
\end{figure}

In Fig.~\ref{PESLv}, we present the PES of $^{296}$Lv projected onto the $(\beta_{20},\beta_{22})$ plane. Among the various trajectories, we selected two for in-depth analysis. The first trajectory, marked in red, encompasses the triaxial shapes, while the second trajectory, depicted in blue, corresponds to the axial shape.
The energy landscape is determined by minimizing the energy on a five-dimensional grid (\ref{sp1grid}) concerning $\beta_{40}$, $\beta_{60}$, and $\beta_{80}$. The inclusion of quadrupole nonaxial deformation $\beta_{22}$ in (\ref{sp1}) significantly modifies the landscape and plays a crucial role in the depiction of the first SP. As depicted on the map, the impact of this effect is evident in the substantial reduction of the axial barrier by approximately 1 MeV.
However, it is crucial to consider that energy mapping in a multidimensional space poses some challenges. When minimizing certain deformations, the dimensionality reduction often results in a PES composed of disconnected patches corresponding to multiple minima in the discarded dimensions. Hence, the actual SP is identified through the imaginary water flow technique in the total deformation space.

In the initial step, various thermodynamic quantities such as entropy, intrinsic level density, and energy-dependent shell corrections are computed as functions of excitation energy for each deformation along the selected fission path.
Neglecting the kinetic energy for the motion in deformation space for each set of deformation $\beta=(\beta_{20},\beta_{22})$, we have
\begin{equation} \label{eq16b}
U(\beta)+E_{pot}(\beta)=E_{0}\equiv \text{constant},
\end{equation}
where $E_{pot}(\beta)=E_{mac}(\beta)+E_{mic}(\beta)$ is the corresponding potential energy and $U(\beta)$ is the excitation energy.
If damping of microscopic correction with excitation energy is taken into account, the Eq.~\eqref{eq16b} should be replaced with
\begin{equation} \label{eq16c}
U^{*}(\beta)+E_{pot}(\beta,U)=E_{0},
\end{equation}
where $E_{pot}(\beta,U)=E_{mac}(\beta)+E_{mic}(\beta,U^{*})$ represents the potential energy with damped microscopic corrections and $U^{*}(\beta)$ is the corresponding excitation energy. Calculating the level density and entropy at $U^{*}(\beta)$, we consistently take into account the excitation energy-dependent potential as well as the change of thermodynamic quantities due to variation of excitation energy concerning the potential with damped microscopic corrections. The properties of the nucleus at various deformations are calculated at the same value of total energy $E_0$ or, equivalently, at the given excitation energy of the GS: $U_{0} = E_0-E_{mac}(GS)-E_{mic}(GS)$.

In this analysis, the nuclear system is conceptualized as a microcanonical ensemble isolated from energy exchange with its environment. Consequently, the total energy, denoted as $E_{0}$, remains constant during the evolution of the nuclear shape. Given that the excitation energies under consideration are substantially larger than the alterations in the PES with deformation, our computations employ an approximate thermodynamic methodology to ascertain level densities, following the framework proposed by Decowski (1968) \cite{Decowski1968}. This allows us to introduce a local temperatures $T$ corresponding to each deformation set. Microscopic corrections, $\delta F_{shell, pair}(\beta, T)$, and excitation energies, $U(\beta, T)$, are then calculated at each point of PES as elaborated in Sec.~\ref{sec:lev1}.
 Subsequently, the energy-dependent microscopic component of the potential, $E_{mic} (\beta, U) = \delta F_{shell}(\beta, U) + \delta F_{pair}(\beta, U)$, is derived and utilized in Eq.~\eqref{eq16c} to determine $U^{*}(\beta)$ and $E_{mic} (\beta, U^{*})$.

\subsection{Fission Paths}
\begin{figure}[]
\centering
\includegraphics[width=0.5\textwidth] {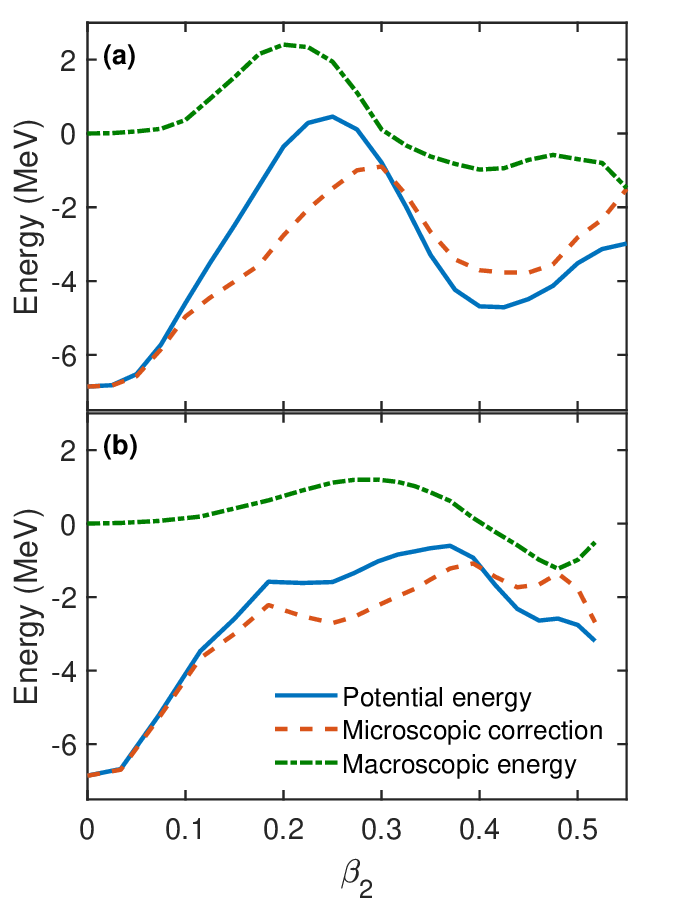}
\caption{The variation of potential energy, macroscopic energy, and microscopic correction at zero excitation energy is depicted as a
the function of the deformation parameter $\beta_{2}$ for axial (a) and triaxial (b) paths of fission of $^{296}$Lv.}
\label{Epot0}
\end{figure}

Our detailed analysis will focus on the entropy and level density parameters along specific fission paths.
We have chosen the paths that exhibit the most significant gradient drop in all variables and have determined the potential to ensure these paths pass through two uniaxial SPs. The axial splitting path is depicted in blue in Fig. \ref{Epot0}.
Additionally, the macroscopic and microscopic components of the potential energy are shown in green and red, respectively.
In Fig. \ref{Epot0}(b), we show the same analysis, maintaining the same color hierarchy but for the nonaxial path. Here, we observe notable differences with Fig. \ref{Epot0}(a). There is essentially no macroscopic barrier along this path. Instead, the shell effect, as expected, creates the barrier and stabilizes the system. Furthermore, we can see that the triaxial barrier is extensive and spread out, whereas the axial barrier is more compact. The competition between the height and shape of these barriers becomes crucial when considering the transmission process, especially when the available energy is lower than the fission barrier (tunneling).

\begin{figure}[]
\centering
\includegraphics[width=0.5\textwidth] {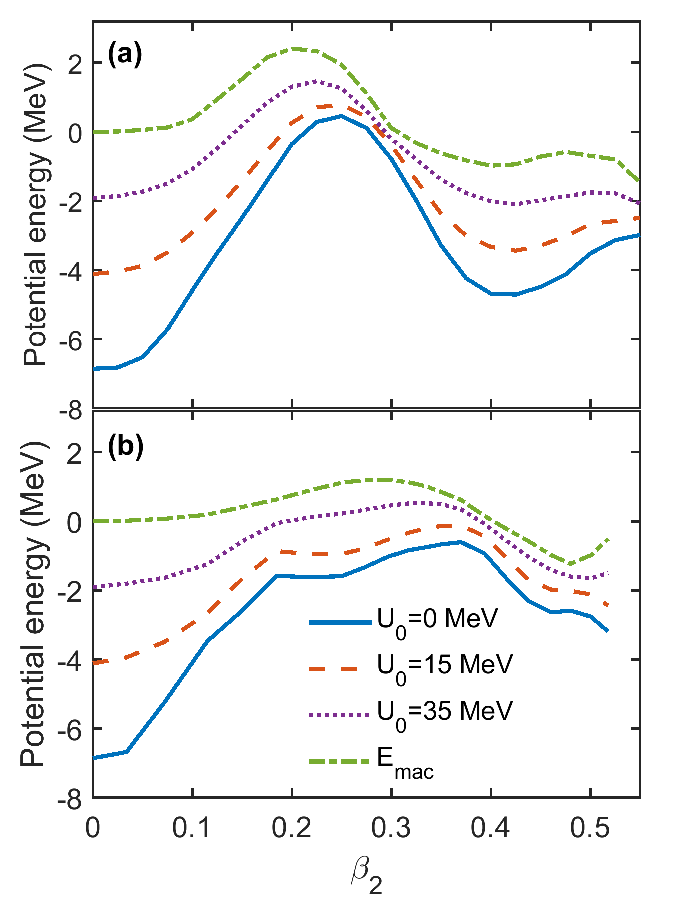}
\caption{The variation of potential energy along axial (a) and triaxial (b) fission path at $U_{0}=0,15,35$ MeV, together with the macroscopic energy. Here, $U_{0}$ is the excitation energy calculated concerning the GS without the damping of shell correction.}
\label{EpotU}
\end{figure}

The influence of excitation energy on the GS and the SP is evident, as shown in Fig \ref{EpotU}. The GS represents the lowest energy configuration, characterized by a well-defined shell structure. As the excitation energy increases, it introduces additional energy into the system, potentially destabilizing the shell structure. On the other hand, the SP corresponds to a highly deformed configuration, where the competition between shell effects and the deformation energy is crucial. The excitation energy has a more indirect impact on the SP, primarily affecting the overall energy of the system rather than the specific internal structure.
Nevertheless, the destructive nature of excitation energy on the entire fission barrier is observed. At higher excitation energies, the barrier gradually diminishes as the influence of shell effects becomes less pronounced and the barrier approaches the macroscopic energy. Eventually, the threshold for fission disappears as the increasing excitation energy washes out the barrier. Also, with increasing $U_0$, the potential energy as a function of $\beta_2$ becomes flat near the GS.

\begin{figure}[]
\centering
\includegraphics[width=0.5\textwidth] {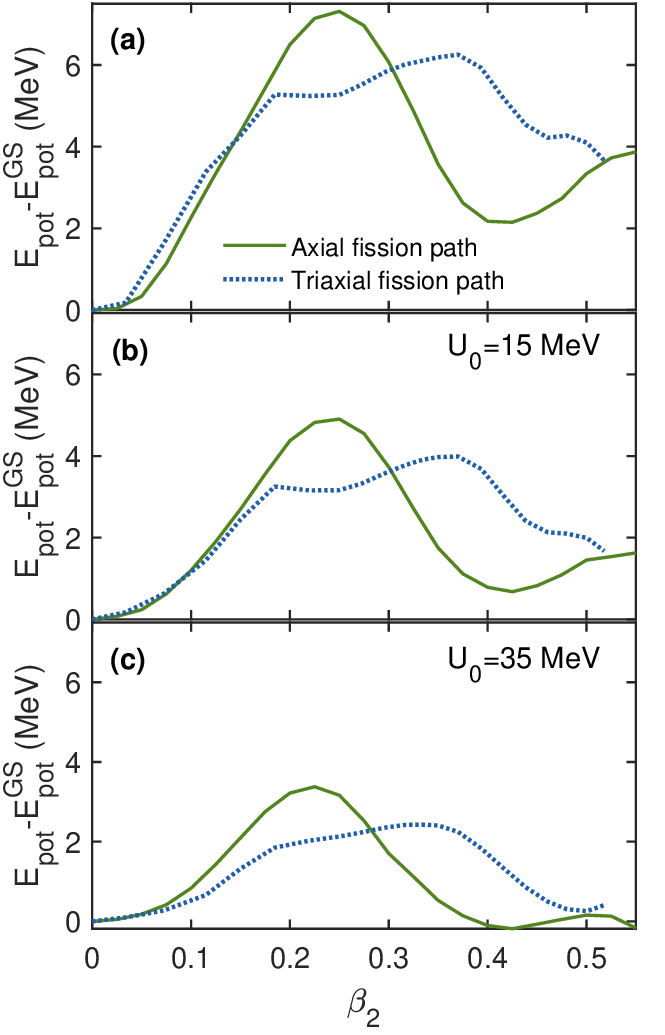}
\caption{The variation of fission barriers with excitation energy $U_{0}=0,15,35$ MeV (a-c), for axial (solid lines) and triaxial (dotted lines) fission path.}
\label{Barrier shape}
\end{figure}

In Fig.~\ref{Barrier shape}, we present the variation of fission barriers $E_{pot}-E_{pot}^{GS}$ with excitation energy for both axial (solid lines) and triaxial (dashed lines) fission paths.
As stated before, $U_{0}$ is the excitation energy concerning cold GS. The actual excitation energy of the GS is smaller than $U_{0}$ because of the damping effects. For instance, at $U_{0}=25$ MeV, the GS excitation energy concerning potential with damped microscopic corrections is $U^{*}(GS)=20.93$ MeV. The damping effect reduces fission barrier corresponding to triaxial(axial) path from $B_{f}^{tri(ax)}=6.25(7.31)$ MeV for cold nucleus to $B_{f}^{tri(ax)}=3.05(3.93)$ MeV at $U_{0}=25$ MeV. Also, the damping effect shifts the triaxial SP from $\beta_{2}=0.37$ to $\beta_{2}=0.33$, and the axial SP from $\beta_{2}=0.25$ to $\beta_{2}=0.22$.

\subsection{Level-density parameter}

\begin{figure}[]
\centering
\includegraphics[width=0.5\textwidth] {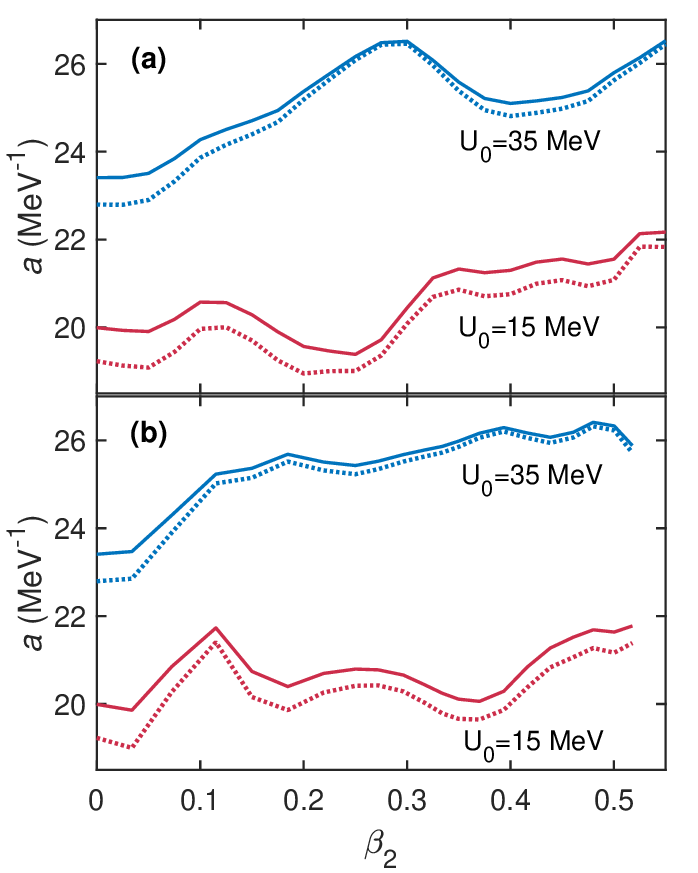}
\caption{(a): Level-density parameters obtained for axial fission path at $U_{0}=15,35$ MeV with (dotted lines) and without (solid lines) consideration of damping of microscopical effects. (b): The same as panel (a), but for the triaxial fission path.}
\label{LDP}
\end{figure}

This work explores variations of level-density parameters with deformation along the fission path. The level-density parameter at the GS corresponds to particle emission channels such as neutron emission ($a_{n}$), while its value at the SP corresponds to fission ($a_{f}$). It is widely recognized that even the most minor change in the values of $a_n$ and $a_f$ can significantly impact the accuracy of calculations of the nuclear survival probability. Our investigation encompasses the entire cleavage path, both axially (Fig. \ref{LDP}(a)) and non-axially (Fig. \ref{LDP}(b)) symmetric at $U_0=15$ and $35$ MeV. The choice of pathway significantly affects the behavior of the level-density parameter. The variability is more pronounced along the axial path and appears smoother along the nonaxial path. Additionally, the results demonstrate that the intricate topology of the PES has a more significant impact on the parameter at lower energies than anticipated.

As evident from Fig. \ref{LDP} (see dotted lines), the damping of microscopic correction systematically influences both fission pathways. Although this effect may seem small, it is not negligible because it affects the GS and the SP differently. Therefore, the damping significantly impacts $a_{f}/a_{n}$ ratio.
It is interesting to note that at small $U_{0}$, the value of $a(SP)$ is comparable or even smaller than $a(GS))$. This is in line with the conclusion of Ref.~\cite{Azam_ratios_2022} that the ratio $a_{f}/a_{n}$ exhibits an increasing trend with excitation energy approaching its asymptotic value less than 1.1.

\begin{figure}[]
\centering
\includegraphics[width=0.5\textwidth] {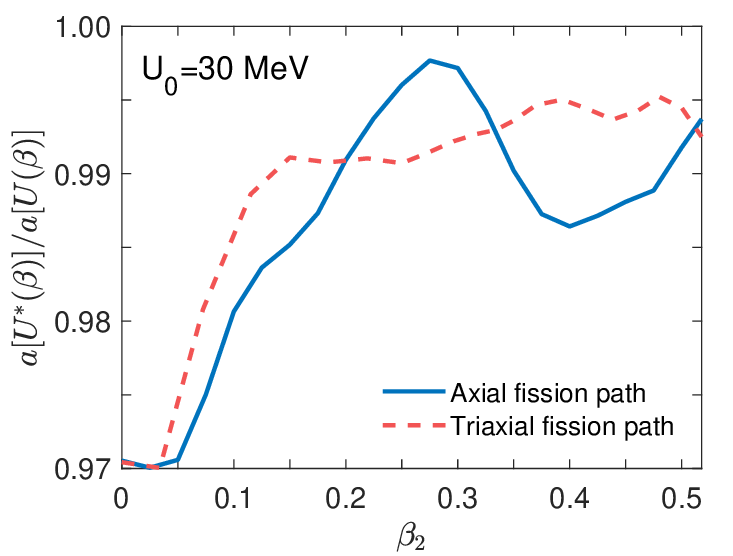}
\caption{The ratio of level-density parameters obtained with account for damping of microscopic corrections to those obtained without account for damping of microscopic corrections along axial (solid line) and triaxial (dashed line) fission path at $U_{0}=30$ MeV.}
\label{LDPratio}
\end{figure}

To highlight the damping effect more prominently,
the ratio of level-density parameters calculated with and without damping along axial (solid line) and triaxial (dotted line) fission paths is presented in Fig. \ref{LDPratio}. The calculations are performed at  $U_{0}=30$ MeV. The most significant impact occurs near the GS.
This effect can be attributed to the GS's significant shell correction and large excitation energy. A similar but less pronounced effect is observed for the axial path at a deformed minimum around $\beta_{20}\sim 0.4$. It is noteworthy that the observed effect of damping never exceeds 5$\%$.

\subsection{Entropy}

\begin{figure}[]
\centering
\includegraphics[width=0.5\textwidth] {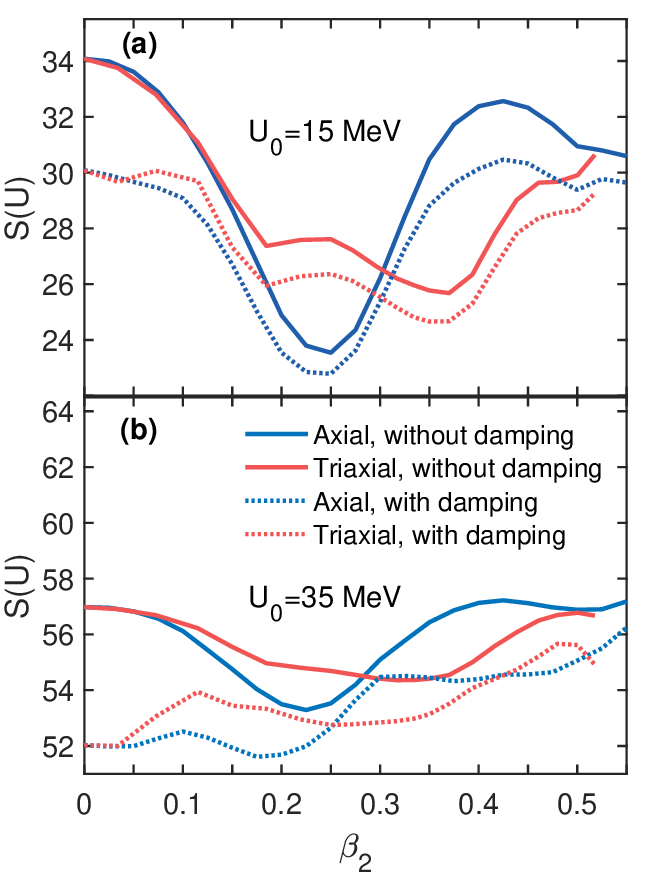}
\caption{Comparison between entropy along axial (blue lines) and triaxial fission paths (red lines) at excitation energies $U_{0} = 15$ and $35$ MeV with (dotted lines), and without (solid lines) consideration of the damping of microscopic effects.}
\label{EntropyU}
\end{figure}

Figure \ref{EntropyU} displays two panels illustrating the variations in entropy along the axial and nonaxial fission paths at energies $U_{0}=15$ and $35$ MeV. Dotted lines depict the entropy values calculated with damped microscopic corrections in potential, while the solid lines represent the results without accounting for damping. By comparing these two sets of curves, we can assess the influence of damping on the entropy along these paths. Notably, regardless of the chosen path, the entropy systematically decreases when potential damping is included.

It is apparent that at low energies, the behavior of entropy, to a large extent, is determined by the intricacies of the potential energy and of the density of single-particle states along the path. The nucleus undergoes significant deformation changes as it moves from the GS to the SPs. Along these paths, the elongation and stretching of the nucleus lead to an increase in the density of single-particle states. Thus, the entropy of the deformed configurations is enhanced. However, at low energies, the impact of excitation energy on entropy is dominant, so the maximum entropy is observed at the GS. With the increase of excitation energy, the trend of entropy curves is still defined by the shape of potential energy. Still, it is less pronounced because of the growing role of the density of single-particle states.

The account for the damping completely changes the behavior of entropy. The damping of microscopic correction suppresses the role of excitation energy in entropy but leaves the role of the density of single-particle states untouched. As seen from Fig.~\ref{EpotU}, because of the significant damping of microscopic corrections at the GS, the excitation energy of the GS decreases more in comparison with the deformed configurations. Therefore, the gain in entropy of the GS due to the more considerable excitation energy is reduced. This leads to the smearing of the entropy as a function of deformation, irrespective of whether triaxiality is considered.

At larger $U_{0}$, the entropy curve gets flattened around the GS due to the significant reduction of the excitation energy for the configurations around the GS and the more substantial density of states for deformed configurations. Moreover, the GS's entropy becomes smaller than that of SPs.

It is worth noting that the difference between entropies at the axial and triaxial SPs decreases regardless of whether the damping effect is considered.

\begin{figure}[]
\centering
\includegraphics[width=0.5\textwidth] {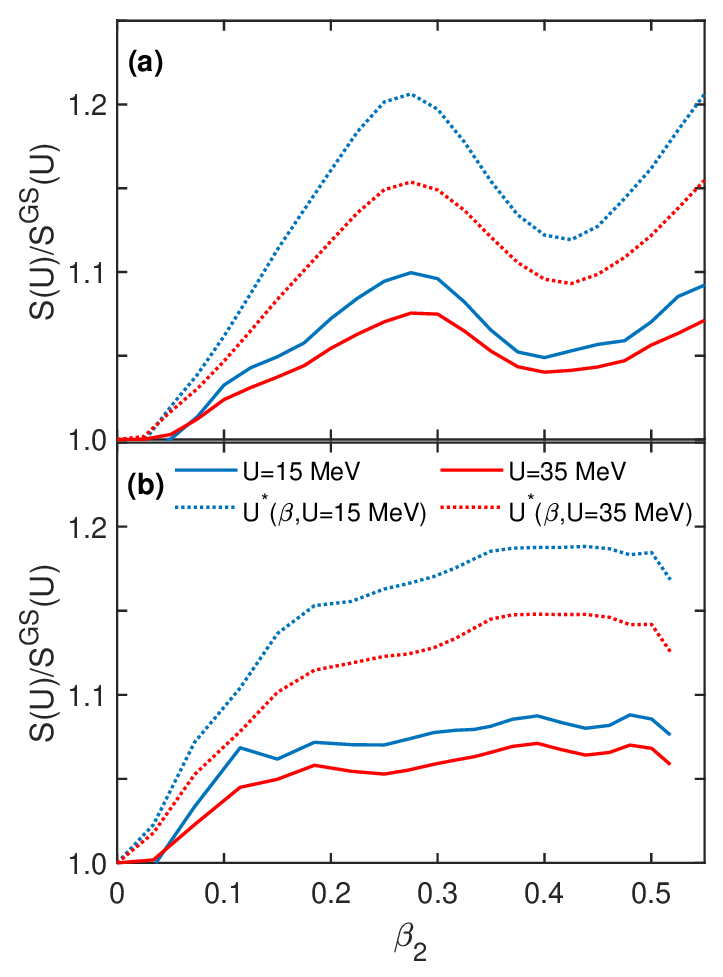}
\caption{(a) The ratios of entropies calculated at each deformation at fixed excitation energies $U=15$ and 35 MeV to those of the GS with considering potentials without damping  (solid lines) for axial fission path. The calculation at excitation energies concerning the potentials with damped microscopic corrections $U^{*}$ are presented with dotted lines. (b) the same as panel (a), but for the triaxial path.}
\label{StructureS}
\end{figure}

In Fig.\ref{EntropyU}, both energy and deformation effects contribute to the displayed entropy curves. We performed the calculations with fixed excitation energy along the path to elucidate the effect of deformation on entropy. Note that the assumption of constant excitation energy is not realistic, and we do it only to study the impact of deformation on entropy. In Fig.~\ref{StructureS}, the excitation energies are fixed as $U=15$ and $35$ MeV concerning potentials with undamped microscopic corrections (solid lines). Panel (a) shows the ratio of entropies calculated at each deformation along the axial fission path to those at the GS. Panel (b) shows the same as panel (a) but for the triaxial path.
The trend of these curves resembles the shape of the potential energy. The effect of structure gradually decreases with excitation energy. It is seen in Fig. \ref{StructureS} that at a given excitation energy, the behavior of entropy is determined by the increased complexity of the deformed configurations. Entropy is generally higher at the SP than at the GS because the SP represents a more disordered and diverse configuration of nucleons and benefits from using multiple energy shells.
The calculations at $U^{*}$ values corresponding to the considered energies are shown with dotted lines. As seen,  the structure effect is more potent in this case. The damping at the GS is more significant than that at larger deformations. Hence, the enhancement of the entropy ratios at the SP seen from the figure is both due to more significant state density and slower damping of microscopic corrections.

\subsection{Fission probability}

To investigate the impact of entropy on decay properties, we employ a simplified approach to calculate the decay rates for two distinct paths. Our analysis involves discretizing the trajectories into a finite number of grid points: $i=(1:n_{\text{max}})$ for the axial path and $i=(-n_{\text{max}}:-1)$ for the triaxial path. The grid point $i=0$ corresponds to the configuration of the GS. Each point is characterized by a set of deformation parameters
$\beta_i=(\beta^{(i)}_{20},\beta^{(i)}_{22})$.

\begin{figure}[]
\centering
\includegraphics[width=0.5\textwidth] {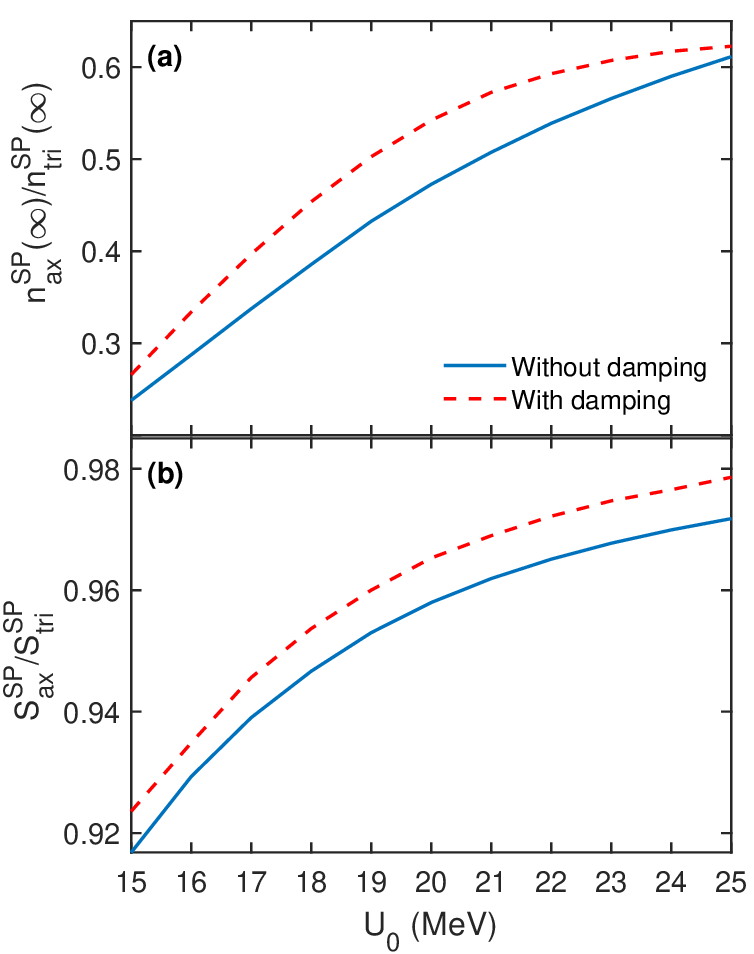}
\caption{Panel (a): The ratio of fission probability through the SP of axial fission path to that of triaxial fission path as a function of excitation energy. Panel (b): The entropy ratio at the SP of the axial fission path to that of the triaxial fission path.  The solid blue lines and dashed red lines indicate the calculations with and without damping of microscopic corrections in potential, respectively.}
\label{FissionProbRatio}
\end{figure}

The probability $n(\beta_i)$ of the nuclear system being populated at time $t$ in a state characterized by deformation $\beta_i$ is determined by a set of master equations \cite{Moretto1975}
\begin{equation} \label{M1}
\frac{d n(\beta_{i})}{dt}=  \Lambda_{i,i+1}n(\beta_{i+1})+\Lambda_{i,i- 1}n(\beta_{i-1})-\Lambda_{i\pm 1,i}n(\beta_i).
\end{equation}
We assume that only neighboring systems ($i,i\pm1$)
 are connected. Consequently, within our simplified model, systems with different trajectories (axial or triaxial) are only linked through the GS. This approximation seems appropriate at low excitation energies. In Eq.~\eqref{M1}, the symbol $\Lambda_{i,f}$, ($f=i\pm1$) represents the transition rates, which can be expressed in terms of microscopic transition probabilities and the level density of the final states.
 Using equal grid intervals and assuming the same transition strength $\lambda_0$ for both axial and triaxial paths, we calculate the transition width as
\begin{eqnarray} \label{M2}
&&\Lambda_{i,f}=\lambda_{i,f}\rho(\beta_{f}),\nonumber \\
&&\lambda_{i,f}=\lambda_{f,i}=\frac{\lambda_{0}}{\sqrt{\rho(\beta_i)\rho(\beta_f)}}.
\end{eqnarray}
We assume that once the system reaches any SPs, it certainly undergoes fission.
Since the system inevitably undergoes fission, the fission probabilities through the SP of axial [$n^{SP}_{ax}(t)$] and triaxial [$n^{SP}_{tri}(t)$] paths are connected as $n^{SP}_{ax}(\infty)+n^{SP}_{tri}(\infty)=1$. The ratios of fission probabilities, corresponding to an axial and triaxial paths calculated with and without accounting for damping effects, are displayed in Fig.~\ref{FissionProbRatio} (a) as a function of $U_{0}$. The ratio of entropy at the SP of the axial fission path to that of triaxial fission path $S^{SP}_{ax}/S^{SP}_{tri}$ is also displayed in the panel (b).
The contribution of the considered paths to total fission depends on their relative entropies at the SP.
The triaxial path exhibits dominance at lower energies, while at higher energies, the contributions of both the axial and triaxial paths become comparable. As seen from Fig.~\ref{FissionProbRatio} (b), the $S^{SP}_{ax}/S^{SP}_{tri}$ ratios both for the calculations with and without damping of microscopic corrections in potential are approaching asymptotic values after $U_{0} = 25$ MeV. Consequently, the ratios of fission probabilities are expected to exhibit a constant behavior at $U_{0}>25$ MeV.
However, it is worth noting that when the damping effect is taken into account, the entropy values from the GS to the SP get closer to each other with increasing excitation energy so that, as shown in Fig.~\ref{EntropyU} (b), at $U>25$ MeV they become almost constant. Hence, the assumption of two distinct paths for evaluating fission probability becomes inadequate. This finding suggests that to model the fission process accurately, particularly when considering the damping effect, the competition between the axial and triaxial paths should be accounted for all intermediate points along the paths rather than solely at the GS.

Without damping, the GS  entropy remains higher than at the SP, even at higher excitation energies.
This implies that there is a probability for the system to return to the GS and subsequently have the opportunity to choose another path.
However, when the damping effect is considered, because of the relatively small difference between SP and GS entropies, it progresses toward the SP once the system selects a path. It does not revert to the GS.
With damping microscopic corrections in potential and decreasing the variation of entropy along the fission paths, the axial path, which has an SP closer to the GS, achieves a non-zero decay probability more rapidly than the triaxial path. This effect is represented in decay constants $\lambda$ which are determined by fitting the $n^{SP}(t)$ with the following form
\begin{equation}
n^{SP}(t)=n^{SP}(\infty)(1-e^{-\lambda t}).\nonumber
\end{equation}

\begin{figure}[]
\centering
\includegraphics[width=0.5\textwidth] {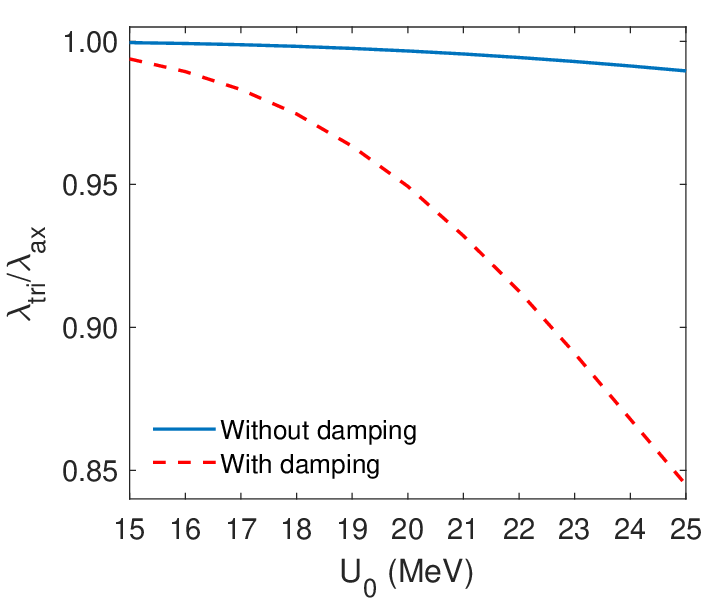}
\caption{The ratio of the decay constant corresponding to the triaxial path to that of the axial path is shown as a function of excitation energy. The solid and dashed lines indicate the calculations with and without damping of microscopic corrections, respectively.}
\label{DecayFigs3}
\end{figure}

The ratio  $\lambda_{tri}/\lambda_{ax}$  of decay constants corresponding to the triaxial and axial pathways is displayed in Fig.~\ref{DecayFigs3}. As shown, with an account of the damping effect, the axial path reaches its asymptotic probability $n^{SP}_{ax}(\infty)$ faster than the triaxial path, and this effect increases with excitation energy.

In the current research, we must acknowledge that collective effects have not been incorporated into our analysis. Our focus has been primarily on examining intrinsic level densities along fission pathways under the presumption of a gradual and slow motion in collective degrees of freedom. It is evident that collective phenomena, particularly rotational enhancement, have the potential to influence this process through the lowering of states associated with triaxial rotation. However, it is important to note that the qualitative trends observed in our results as a function of excitation energy remain consistent, irrespective of whether the rotational effects are included. For instance, the ratios of $n_{ax}^{SP}(\infty)/n_{tri}^{SP}(\infty)$, as depicted in Fig.~\ref{FissionProbRatio}, exhibit a similar pattern even when collective effects are considered, though there is a noticeable downward shift by approximately 10-15 percent.
This observation underscores the robustness of our findings, suggesting that while the inclusion of collective effects leads to quantitative adjustments, the fundamental trends and conclusions drawn from our study are preserved.

\section{\label{sec:level14}CONCLUSIONS}
Our analysis emphasizes that in the study of fission, it is essential to consider not only the shape of the PES but also the entropy,
which incorporates both energy and structural effects.
As demonstrated for the superheavy nucleus $^{296}$Lv, variations in structural effects at different points of the fission path can lead to significant changes
of the level density and, correspondingly, of the entropy, especially in the case of axial symmetry.
Furthermore, our work highlights the significant influence of suppressing shell effects in certain regions where it is strong and cannot be neglected.
At sufficiently high excitation energies, the fission process exhibits iso-entropic behavior.
While at lower energies, the entropy demonstrates more pronounced variations.
By employing the master equation, we computed the decay probabilities that vary with time, considering the permissible symmetry in the fission pathway.
As shown, the behavior of entropy along the fission paths influences the time-dependent fission probability. The decay constants corresponding to the axial and triaxial paths are comparable.
To provide a more precise analysis,
fission calculations should be conducted on a multi-dimensional PES, considering different trajectories leading to the SP region.
However, to gain a comprehensive understanding, one should expand the study to encompass additional degrees of freedom and allow the system to select an optimal path. This extension will provide a more thorough exploration of the fission process and its dependence on entropy.

\section*{ACKNOWLEDGEMENTS}

M. K. was co-financed by the COPIGAL project.
T.M.S, G.G.A., and N.V.A. were supported by the Ministry of Science and Higher Education of the Russian Federation (Moscow, Contract No. 075-10-2020-117).

\end{document}